\documentclass{iaus}
\usepackage{graphicx}
\usepackage{url}

\title[LOFAR] %% give here short title %%
{The scientific potential of LOFAR for time-domain astronomy}

\author[Rob Fender]   %% give here short author list %%
{Rob Fender on behalf of the LOFAR Transients Key Science Project}
\affiliation{Physics \& Astronomy, University of Southampton, SO17 1BJ, UK
\\ email: {\tt r.fender@soton.ac.uk} \\
For TKSP member list see: {\bf www.transientskp.org/team.shtml}\\}

\pubyear{2011}
\volume{285}  %% insert here IAU Symposium No.
\pagerange{--}
\jname{New Horizons in Time Domain Astronomy}
\editors{E. Griffin, R. Hanisch, R. Seaman (eds).}
\begin{document}

\maketitle

\begin{abstract}
LOFAR is a groundbreaking low-frequency radio telescope currently
nearing completion across northern europe. As a software telescope
with no moving parts, enormous fields of view and multi-beaming, it
has fantastic potential for the exploration of the time-variable
universe. In this brief paper I outline LOFAR's capabilities, as well
as our plans to use it for a range of transient searches and some
crude estimated rates of transient detections.

\keywords{accretion, stars:binaries, stars:pulsars, stars:supernovae:general, ISM: jets and outflows, radio continuum: general}
\end{abstract}

\firstsection % if your document starts with a section,
\section{Introduction}

\begin{figure}[t]
\begin{center}
\includegraphics[width=11.0cm]{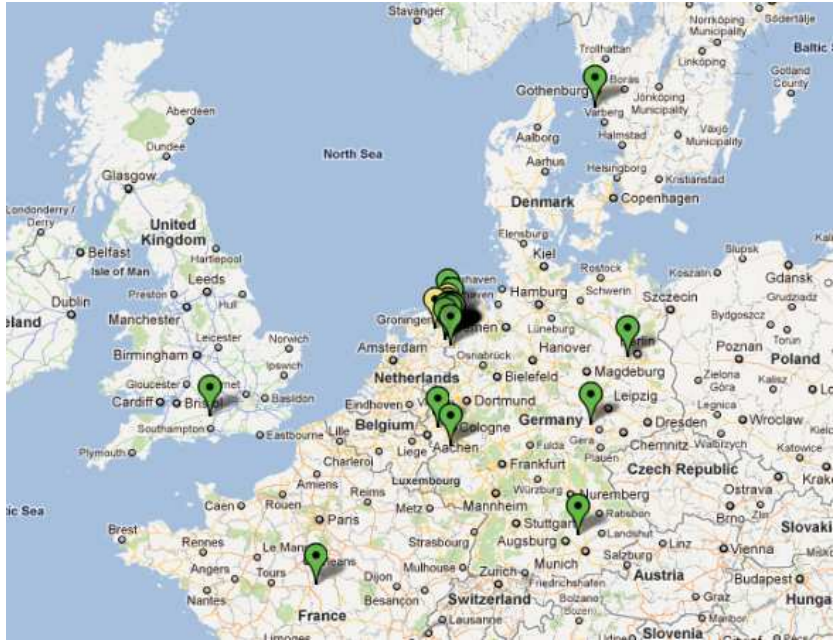} 
\caption{The distribution of complete LOFAR stations across Europe.}
\label{europe}
\end{center}
\end{figure}

LOFAR, the Low Frequency Array, is a large, low-frequency radio
telescope in northern Europe, led by ASTRON. Construction of the
array, which has its core collecting area in The Netherlands, with
international stations in France, Germany, Sweden and The UK, is
nearly complete, and astronomically interesting data are now being
taken. LOFAR operates in the 30--80 and 120--240 MHz frequency ranges. The
80--120 MHz frequency gap corresponds to the FM radio bands at which
frequencies astronomical observations would be
impossible\footnote{Unless northern europe could be persuaded to stop
  night-time FM radio broadcasts for a few weeks in the interests of
  finding the {\em Epoch of Reionisation} signal..}. Construction of
the array is almost complete -- See Fig \ref{europe} for the
distribution of operating LOFAR stations across Europe.  In addition,
observations are occasionally possible to frequencies as low as 15
MHz. For a full reference paper on LOFAR see van Haarlem et
al. (2012).

LOFAR has six {\em key science projects} (KSPs), one of which is {\em
  Transients} (principal investigators Fender, Stappers, Wijers).  The
remit of the TKSP covers essentially all transient and variable
astrophysics, including commensal searches of all data (ultimately in
near-real-time, although this functionality is not yet implemented).
The TKSP covers both time-series and image-plane searches for
transients and variables, including pulsars (Stappers et al. 2011).
The adoption of transients and variables as key science drivers for
the project is a theme for most of the large SKA pathfinders and
precursors, in general unlike older radio facilities. However,
time-series and image-plane transients have been separated for both
ASKAP (which has {\em CRAFT} and {\em VAST} respectively) and MeerKAT
({\em TRAPUM} and {\em ThunderKAT}). This makes some sense from a
techniques point of view, although there is some overlap in the
science.

\section{Types of radio transients}

So what are these transients which we're looking for? In planning for
transients searches, whether via targeted surveys or blind commensal
studies, we can crudely separate events by both astrophysics and
techniques.

Incoherent synchrotron emission arises from the acceleration of
relativistic electrons in magnetic fields. This has a brightness
temperature (intensity) limit of $\sim 10^{12}$K, and is associated
with essentially all events which inject kinetic energy into the
ambient medium (e.g. jets of all types, nova and supernova
explosions). These events tend to evolve relatively slowly, especially
at low frequencies (where the optical depth is higher), and so with
LOFAR we do not expect to see variability on timescales shorter than
the standard imaging timescale of 1--10 sec.

Coherent events, such as radio pulsars or masers, can have much higher
brightness temperatures. Like synchrotron events they can be at times
associated with extreme astrophysical environments. The much higher
brightness temperatures means that the variability timescales can be
much shorter, and even astrophysically distant objects can vary on
timescales much shorter than the standard imaging timescale.  However,
such short bursts are dispersed and scattered in the intervening
interstellar medium, and this needs to be corrected for in studying
the intrinsic properties of the burst (this is a well-understood
problem in the field of radio pulsars). Note that there can be
considerable overlap between these two sets of objects and techniques:
e.g. variability of coherent sources can be detected and tracked in
images.

\begin{figure}[t]
\begin{center}
\includegraphics[width=12.0cm]{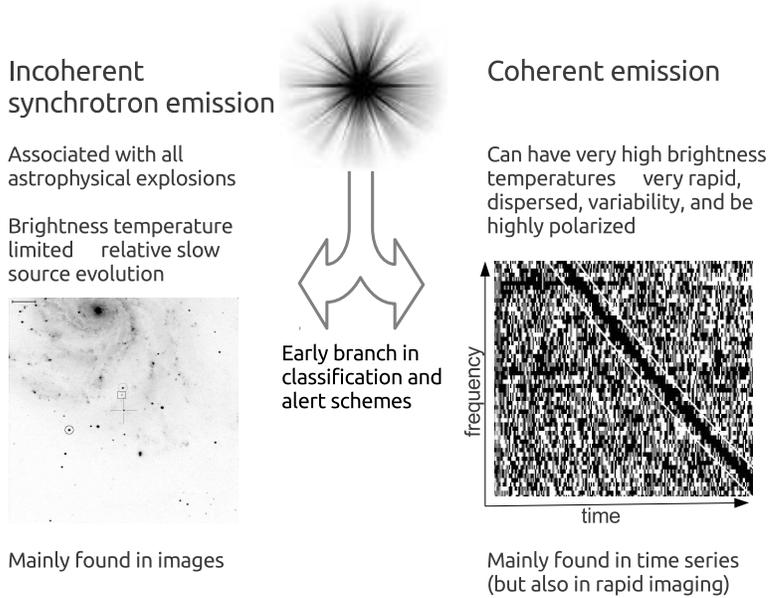}
  \caption{Transients can be divided into incoherent synchrotron and
    coherent evevents, which corresponds roughly to a divide also in
    the techniques used to find them (image plane vs time series).}
   \label{flavours}
\end{center}
\end{figure}

Fig \ref{flavours} summarises this dichotomy in radio transients, and
is indicative of how early attempts at automated classification
pipelines might make an early branch based upon variability timescales
and polarisation characteristics.

\section{Finding transients with LOFAR}

LOFAR can operate in a variety of modes, all of which can be important
for the study of transients and variables. Furthermore, all of these
modes can be operated at a variety of levels from a single station to
the entire pan-European array.

{\bf Interferometric} LOFAR is a `software telescope' with essentially
no moving parts. Pointing of the array and/or individual stations is
done by introducing delays appropriate to a certain direction on the
sky (phased array). Different frequencies can be therefore set to
observe in different directions by introducing different delays. LOFAR
can already, as a standard imaging mode, produce 8 beams each of 6 MHz
bandwidth. In the low band these beams can be placed anywhere on the
sky; in the high band they are limited by the beam of the high band
tiles, which each have an analogue beamformer. This allows for an
extraordinary instantaneous field of view: $8 \times 90 = 720$ square
degrees in the low band, and $8 \times 25 = 200$ square degrees in the
high band. In other words, in the low band the entire northern
hemisphere could be mapped in less than 30 sets of pointings (with
sparse tiling). Initial processing of wide-field surveys for
transients, including the MSSS survey, due for late-2011 through
early-2012, will only localise sources to a few arcmin, but later
and/or responsive observations could localise interesting sources
(including transients) to arcsec precision.

{\bf Timing} LOFAR also has high time resolution (`pulsar') modes,
which can achieve 10s of ns time resolution and map either full field
of view with sensitivity $s \propto N^{-1/2}$ (incoherent sum) or the
synthesised beam with sensitivity $s \propto N^{-1}$ (where smaller
$s$ is better). Recently it has been possible to simultaneously record
data from over 100 coherent tied-array beams and tile out the entire
HBA field. See Stappers et al. (2011) for more on searches for fast
transients with LOFAR.

{\bf Direct storage}
The LOFAR Transients Buffer Boards (TBBs) can be used to record up to
several seconds of full bandwidth antenna level data (or longer, in a
trade-off with bandwidth), {\em before} the beam-forming stage. This
means that beams can be formed retrospectively in a certain direction
anywhere in the sky (LBA) or tile beam (HBA) upon receipt of an
`internal' (from LOFAR itself) or `external' (e.g. VOEvent)
alert. Currently this mode is being developed and pursued by the {\em
  Cosmic Rays} KSP (PI Falcke).

\begin{figure}[b]
\begin{center}
\includegraphics[width=11.0cm]{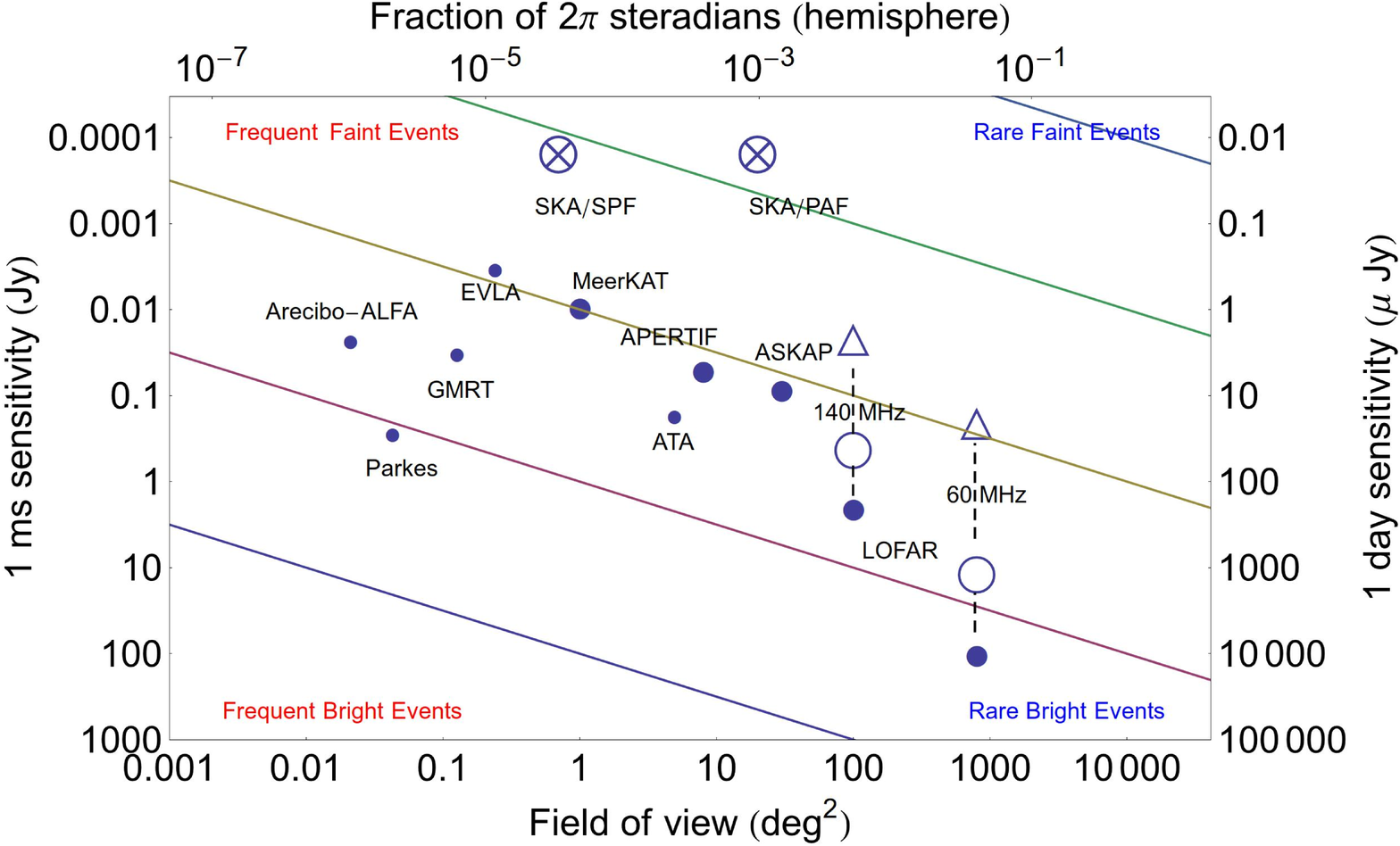}
  \caption{A comparison of sensitivity vs. field of view for a range
    of existing and planned radio telescopes. The solid lines
    represent a constant survey figure of merit (FoM, $\propto \Omega
    s^{-2}$ where $\Omega$ is the field of view and $s$ the
    sensitivity, where smaller $s$ is more sensitive). For LOFAR, the
    points indicate the raw sensitivities, the open circles represent
    a spectral correction for a spectral index of -0.7 (where spectral
    index $\alpha$ is in the sense that $S_{\nu} \propto
    \nu^{\alpha}$), appropriate for optically thin synchrotron
    emission. The open circles correspond to a correction for a
    spectral index of $-2.0$, corresponding to the steepest (aged)
    synchrotron sources, as well as some coherent radio sources (such
    as pulsars and other flavours of neutron star).}
   \label{newfov}
\end{center}
\end{figure}

\section{LOFAR in a global context}

As noted above, LOFAR, ASKAP and MeerKAT have all embraced the science
of radio transients as part of their key science programme. To this
list we can hopefully add APERTIF, the focal-plane array upgrade to
WSRT, to which several transients-oriented proposals have been
submitted as statements of interest for its KSP programme.

In a global context, LOFAR has the widest field of view of any of the
major facilities and, although its suffers in terms of raw mJy
sensitivity compared to e.g. EVLA, MeerKAT, when a spectral correction
is made it can be demonstrated to be a very powerful facility. This is
illustrated by Fig \ref{newfov} in sensitivity and field of view are
compared for a range of world-class radio facilities. The solid
diagonal lines correpond to constant figures of merit (FoM, defined as
$FoM \propto \Omega s^{-2}$, where $\Omega$ is the field of view and
$s$ is the sensitivity, with smaller $s$ more sensitive). The four new
or upgraded GHz facilities, EVLA, MeerKAT, APERTIF and ASKAP, all have
a comparable FoM. For LOFAR it is not until a spectral correction is
made that its survey power becomes apparent (see figure caption for
more details).

\begin{figure}[b]
\begin{center}
\includegraphics[width=11.0cm]{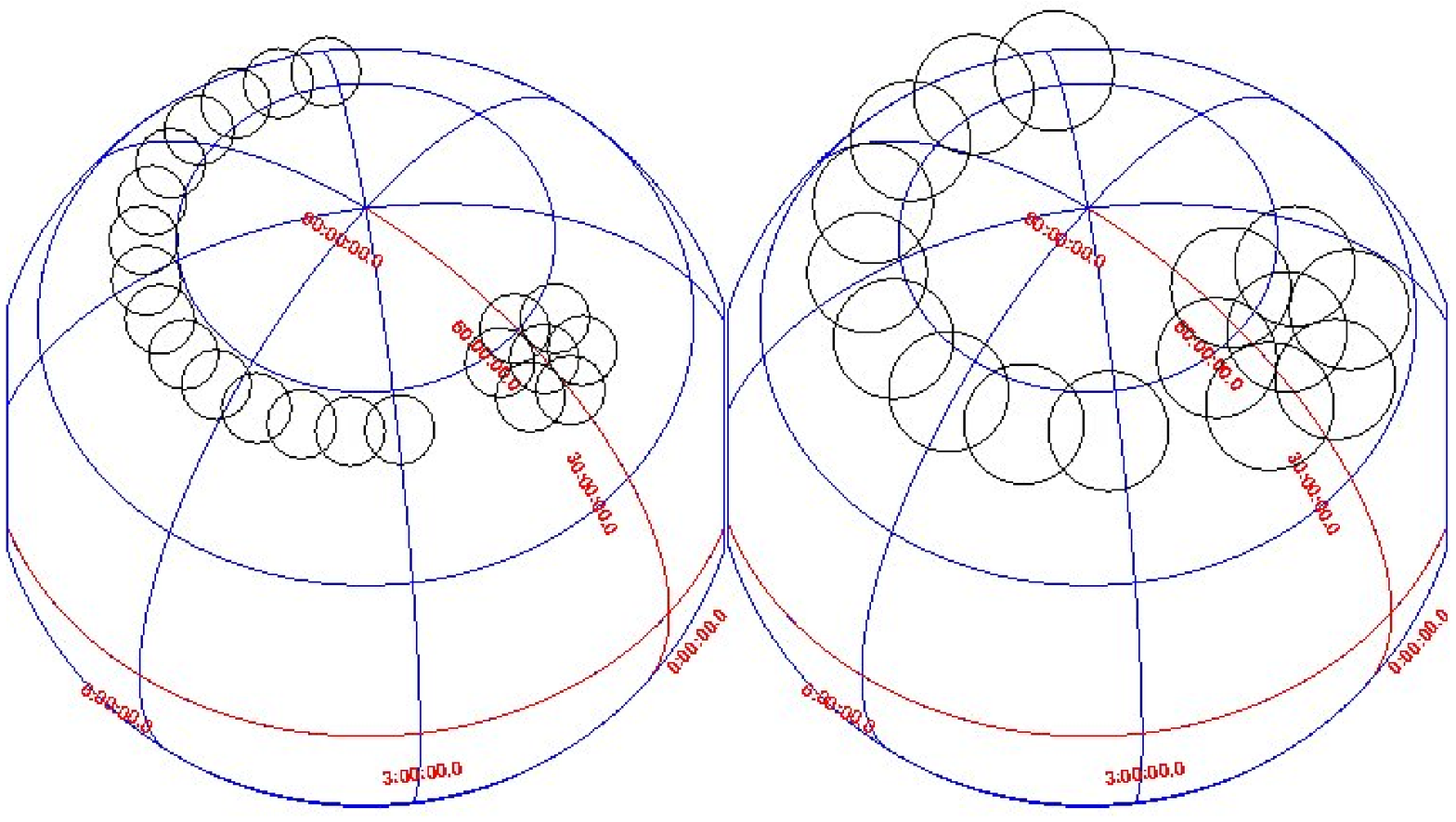}
  \caption{The proposed zenith monitoring programme for LOFAR. During
    intense periods of monitoring, some 2000 square degrees of the sky,
    centred on Dec +54, could be imaged to $\sim$mJy sensitivity every
    24 hr. The figure demonstrates the enormous field of view of LOFAR.}
   \label{zenith}
\end{center}
\end{figure}

\section{Proposed approach}

The LOFAR TKSP have considered several approaches to the detection of
transients and variables, consisting of a programme of targeted
searches, wide-field blind surveys, and commensal searches for
transients.

{\bf Targeted surveys.}
Within the TKSP there are a number of targeted observations and
surveys planned and, in some cases, under way. These include
observations of well-known high-energy astrophysics sources with
variable radio counterparts (e.g. the binary SS 433, the pulsar PSR
0329+54, the blazar PKS 1510-089). They also include targets beyond
the usual suspects, which may provide breakthroughs in some research
areas (e.g. a search of nearby stars for radio bursts from `hot
Jupiters').  In addition, it is our intention to probe transient
parameter space by searching for concentrations of mass and/or exotic
objects at a range of distances (and hence luminosities) from
us. These include globular clusters within our own galaxy, to M31, the
core of the Virgo cluster and beyond.

{\bf Wide-field blind surveys.}
An example of a wide-field blind survey planned for LOFAR is the
zenith monitoring programme, a key component of the `Radio Sky
Monitor' (RSM) programme.  Fig \ref{zenith} illustrates how a small
number of 7-pointing tiles could cover the entire zenith strip (Dec
+54 for LOFAR), with maximum sensitivity (which peaks at the zenith
for the dipoles). For example, with the LBAs at around 60 MHz, a
10-degree wide strip could be covered in 16 tiles, surveying around
2000 square degrees (10\% of the northern hemisphere; note that this
is for fairly dense tiling). Assuming a dedicated phase of RSM observing
(perhaps likely to happen once or twice per year for a few weeks, in
coordination with other multiwavelength facilities) with 100\% of
resources spent on this, this would correspond to $\sim 1.5$ hr on
each field, resulting in an expected r.m.s. of a few mJy. At the
lowest frequencies it only takes two more tiles to cover the entire
northern galactic plane, which may well sample a different population
of transients. At present, we have been monitoring a single part of
the zenith field, containing the bright pulsar PSR 0329+54, and using
that to constrain and estimate the rate of low frequency transients
(Bell, 2011).
Note that the {\em AARTFAAC} project (PI Wijers) plans to use the
LOFAR core in parallel with other observing programmes to perform a
quasi-continuous low angular resolution survey of the entire northern
hemisphere for transients.

{\bf Commensal surveys.}
Since the initial design of LOFAR it has been a stated goal of the
TKSP that LOFAR data are imaged on short ($\leq 10$ sec) timescales in
close to real time, to look for bright transients and variables. This
functionality is yet to be implemented, but is in the design and
commissioning plan for 2012. In the meantime, the TKSP has the goal of
searching all other observations performed by LOFAR in order to look
for transients. This will naturally give us a range of wide-field and
deep surveys (consider e.g. those planned by the {\em Surveys} and
{\em Epoch of Reionisation} KSPs) with which to probe transient phase
space. It is not unreasonable to assume that the entire northern
hemisphere will effectively be surveyed several times per year, as
well as deeper pointings with much higher cadence. 

\section{Predicted rates}

Until recently, the best estimates for predicted rates were around 0.1
sources deg$^{-1}$ for two-epoch transients and a flux density limit
of $\sim$ mJy (Bell 2011, based on Bower et al. 2007). However, Frail
et al. (2011) have noted that this may be an overestimate by up to an
order of magnitude. For the zenith monitoring programme outlined
above, we could have 2000 deg$^2$ surveyed to approximately this
r.m.s. per 24 hr, resulting in a transient detection rate -- from this
programme alone -- of between 20 (Frail) and 200 (Bower) events {\em
  per day}. Considering commensal searches of all data, which may on
any given day be going wide, deep, or some combination, we should
conservatively expect in excess of ten high-significance transient
events per day. One major unknown here is what fraction of events will
turn out to be repeaters -- i.e. the true number of distinct
astrophysical variables found may be lower than this.  Note that it is
our stated policy to report these events as widely as possible,
probably vis some version of the VOEvent protocol (e.g. via
skyalert.org).

\section{Summary}

LOFAR has a wide range of diverse capabilities (multi-beaming,
simultaneous timing and imaging modes, splitting of array in large
numbers of individual stations, re-imaging in the past with the TBBs)
which are ideally suited for exploring transient parameter space. Just
as importantly, there is the will to support this science, with a Key
Science Project dedicated to precisely this exploration. Based on
current estimates, it should find many 1000s of interesting radio
transients per year, providing a huge target list for multiwavelength
follow-up, and providing new tests of our ability to automatically
detect, classify and report such events efficiently.

\noindent
LOFAR: {\bf www.lofar.org}\\
LOFAR Transients Key Science Project: {\bf www.transientskp.org}\\

\end{document}